\documentstyle[11pt]{article}
\begin{document}

\centerline{\bf\large FURTHER STUDY ON U(1) GAUGE INVARIANCE RESTORATION
\footnote{This work is partly supported by the National Natural
Science Foundation of China}}

\vspace{1cm}

\centerline{Chao-Hsi Chang$^{1,2}$, Yi-Bing Ding$^{3,4}$,
Xue-Qian Li$^{1,2,5}$,
Sheng-Ping Tong$^{3},$}

\vspace{0.6cm}

\centerline{Jian-Xiong Wang$^6$ Jiu-Fen Zhao$^{5}$ and Jie-Jie Zhu$^{7}$}

\begin{center}

$^1$ CCAST (World Laboratory), P.O.Box 8730, Beijing 100080, China.\\

$^2$ Institute of Theoretical Physics,
 Academia Sinica, P.O.Box 2735, Beijing 100080, China.               \\

$^3$ Physics Department, Graduate School of Academia Sinica,
Beijing, 100039, China.\\

$^4$ Physics Department, University of Milan, INFN, Via Celoria 16, 20133,
Milan, Italy.\\

$^5$ Department of Physics, Nankai University, Tianjin, 300071, China.\\

$^6$ Institute of High Energy Physics, Academia Sinica, Beijing,
100039, China.\\

$^7$ Department of Physics, University of Science and Technology of China,
Hefei 230026, China.

\end{center}

\vspace{1.5cm}

\begin{center}
\begin{minipage}{12cm}

\noindent{\bf abstract}

To further investigate the applicability of the
projection scheme for eliminating the unphysical divergence $s/m_e^2$ due to
U(1) gauge invariance violation,
we study the process $e^-+W^+\rightarrow e^-+\bar t+b$
which possesses advantages of simplicity and clearness. Our study indicates
that the projection scheme can indeed eliminate the unphysical divergence
$s/m_e^2$ caused by the U(1) gauge invariance violation and the scheme can
apply to very high energy region.

\end{minipage}
\end{center}

\vspace{1cm}

\baselineskip 22pt

It is well known that when an unstable particle exists at the s-channel as an
intermediate boson or fermion, one must introduce its width, otherwise the
integration over final states would blow up as long as the energy $\sqrt s$ is
high enough. The commonly adopted form is the Breit-Wigner form which modifies
the denominator of the propagator ($q^2-M^2$) into ($q^2-M^2+iM\Gamma$) where
$q,M,\Gamma$ are the momentum, mass and width of the unstable particle
respectively. In the simplest way, the width $\Gamma$ can take its measured
value. However, if there is a t-channel $\gamma-$propagator in the Feynman
diagram at the same time, the introduction of $\Gamma$ violates the U(1) gauge
invariance. This violation is due to an ill-treatment i.e. an unreasonable
truncation of the perturbative expansion.

This violation would cause divergent terms such as $\log s/m_e^2$ and $s/m_e^2$.
This divergence only occurs near the $e^--$forward scattering, i.e. small
$\theta$ region, where $\theta$ is the angle between momenta of incoming and
outgoing electrons. The small measure at vicinity of $\theta\rightarrow 0$
makes the logarithmic divergence ($\log s/m_e^2$) benign which can only bring up
negligible changes to the calculations, on contraries, the linear term
$s/m_e^2$ is intolerable and causes the theoretical evaluation of the
corresponding cross section unreasonably large by many orders.

Therefore, to obtain physical results, one should restore or at least partly
restore the U(1) gauge invariance, namely eliminate the disastrous divergent
power term $s/m_e^2$. Many suggestions to remedy this problem were discussed in
literatures \cite {Boos,Zop} for the typical process
$e^+e^-\rightarrow e^-+\bar \nu+u+\bar d$ where an s-channel W-boson propagator
exists and its Breit-Wigner form violates the U(1) gauge invariance.
Later Paravassiliou and Pilaftsis
\cite {Para} carried out a systematic study and restored the U(1) gauge
invariance based on the pinch technique. In their work, the width effects are
compensated in principle.

For practical applications, we need some simple schemes. Argyres et al.
introduced lepton and quark loops and the absorptive part of the triangles
compensates the effect of $\Gamma_WM_W$ which violates the gauge invariance
\cite{Arg}.

Instead, we suggested a "projection scheme" which can project out the
troublesome U(1) gauge violation terms which lead to the unphysical
divergence $s/m_e^2$ \cite{Chang}. This scheme is to choose a special gauge,
then in the theoretical calculations the $s/m_e^2$ terms disappear
automatically, in other words, the U(1) gauge invariance is partly restored.
Even though the logarithmic divergence $\log s/m_e^2$ remains, it is benign,
so that one can tolerate
such terms. Moreover, the contribution from the logarithmic divergence
to the cross section is negligible (up to less than 1\%)
and physical results can be obtained in this scheme (see later part of
this work for more details) for very high energies.

For further study the violation and restoration of the U(1) gauge invariance
and probe the applicability of the projection scheme, we investigate the
process $e^-+W^+\rightarrow e^-+t+\bar b$. 
The concerned Feynman diagrams for this process at the tree level
are presented in Fig.1. The process has some obvious advantages.
First, it is a three-body final state case, so the final-state integration
is much simpler than $e^+e^-\rightarrow e^-+\bar{\nu}+u+\bar d$. Thus it is
easier to study the U(1) gauge invariance and restoration and that is the
aim of this work. Secondly, in this case the concerned s-channel W-propagator
is connected to $t\bar b$ which are real particles in the final states, so that
the intermediate W-boson can never be on its shell and the final state
integration is convergent anyway, thus an involvement of $i\Gamma_WM_W$ in
the W-propagator is not needed at all for obtaining accurate results. In other
words, we can be convinced that the results obtained without introducing the
Breit-Wigner modification of the W-propagator are the physical ones (to order
of $O(\alpha_s^0)$. Our purpose of this work is to study the U(1) gauge
invariance, so that we can take the results obtained without the $i\Gamma_WM_W$
as the a standard and then compare the results with $i\Gamma_WM_W$ existing in
the propagator to it for testifying the applicability and accuracy of our scheme.

In the following, we will give detailed discussions on this problem of U(1)
gauge invariance violation and restoration, mainly show how to obtain reasonable
physical results in our scheme.
\\

(1) The U(1) gauge invariance violation in $e^-+W^+\rightarrow e^-+t\bar b$.

In Fig.1.a, if the propagator of the s-channel W-boson is written in the
Breit-Wigner form as
$${1\over p_+^2-M_W^2+i\Gamma_WM_W},$$
we can check the U(1) gauge invariance, i.e. the Ward-identity.

We write the reaction amplitude as
$$M=\bar u_e(p_1)\gamma^{\mu}{-i\over q^2}u(p_2)(T_a+T_b+T_c)_{\mu}+\Delta M,$$
where $(T_a+T_b+T_c)_{\mu}$ correspond to the effective vertex of
the first three Feynman diagrams in
Fig.1, and the fourth one $\Delta M$ is irrelevant to the U(1) problem. Then,
the Ward identity is
\begin{equation}
A\equiv q^{\mu}\cdot (T_a+T_b+T_c)_{\mu}=0,
\end{equation}
where $q^{\mu}$ is the 4-momentum of the virtual t-channel photon.

A straightforward derivation shows
\begin{eqnarray}
\label{vio}
A &=& q^{\mu}\cdot(T_a+T_b+T_c)_{\mu} \nonumber \\
&=& [(p_+^2-p_W^2){1\over p_+^2-M_W^2+i\gamma_W}-1]\bar u_t(p_2)\gamma^{\alpha}
(1-\gamma_5)v_b(p_3),
\end{eqnarray}
where we take $\gamma_W=\Gamma_WM_W$ following the notation of \cite{Arg} and
other symbols are marked in Fig.1 explicitly.

$A$ is not zero unless
$$\gamma_W=0,\;\;\;\;\;\;{\rm or}\;\;\;\;\;\; p_W^2=M_W^2-i\gamma_W'
\;\;{\rm and}\;\; \gamma_W'=\gamma_W.$$
Because the incoming W-boson is a real on shell particle, usually
$$p_W^2=M_W^2$$
is required. The extra $i\gamma_W'$ represents a deviation from its central
value. In next section, we demonstrate that in a stringent way to restore the
U(1) gauge invariance this extra term is necessary
as long as the s-channel W-propagator is modified to the Breit-Wigner form,
by contrast, in Sec.4, we show that with the projection scheme
one does not need to worry about it after all.\\

(2) Restoration of U(1) gauge invariance.

One of the practical way to restore the gauge invariance was given by Argyres
et al. \cite{Arg}. They considered  the contribution from the
radiative correction to the $WW\gamma$ vertex. They proved that the
absorptive part of a triangle consisting of quarks and leptons compensates the
the effects of $i\gamma_W$ which violates the U(1) gauge invariance, then with
this extra contribution, the Ward identity holds again.

In their derivation for $e^+e^-\rightarrow e^-\bar{\nu}u\bar d$, only one
W-boson is at s-channel (Fig.1.a), so that according to the Cutkosky cutting
rule, only one cut exists and it corresponds to the cut-1 in Fig.2. In our
case $e^-+W^+\rightarrow e^-+t+\bar b$, both cut-1 and cut-2 can exit non-zero.
Observation manifests that the cut-1 gives an absorptive part of the triangle
which compensates the imaginary part of the W-boson propagator, i.e.
$i\gamma_W$, whereas the cut-2 is also non-zero and corresponds to $i\gamma_W'$.
Argyres et al. proved \cite{Arg} that the cut 1 fully compensates
of the imaginary part
of the s-channel W-boson $i\gamma_W$, so a non-zero contribution of cut-2
needs to be compensated by other sources, otherwise, the Ward identity would be
violated again. Thus as claimed above, the on-shell condition of the
incoming W-boson would be changed to $p_W^2=M_W^2+i\gamma_W'$. In general,
$\gamma_W$ does not need to be equal to $\gamma_W'$. From eq.(\ref{vio})
one can notice that if we write $p_W^2=M_W^2+i\gamma_W'$ and let $\gamma_W=
\gamma_W'$, the fermion triangle is not needed and the Ward identity holds
automatically. However, as $i\gamma_W\neq i\gamma_W'$, eq.(\ref{vio}) tells us
that the Ward identity is violated at tree level and we can restore it by
taking into account of higher order correction, i.e. the absorptive part of the
fermionic triangle. Obviously, the cut-1 of Fig.2 compensates $i\gamma_W$
whereas, the cut-2 would compensate $i\gamma_W'$.

Moreover, since in process $e^+e^-\rightarrow e^-+\bar{\nu}+u+\bar d$, the
outgoing quark and antiquarks are $u$ and $\bar d$, whose masses are small
and can be neglected, due to the CVC and PCAC theorems, $p_+^{\alpha}
\bar u_u(p_3)\gamma_{\alpha}(1-\gamma_5)v_b(p_4)\approx 0$ where $p_+=p_3+p_4$.
Then the corresponding Lorentz structure for $\Gamma_{\alpha\beta\mu}$ does
not contain terms proportional to $p_+^{\alpha}$. But in our case, for the
vertex $Wt\bar b$, this simple relation does not exist and the Lorentz
structure becomes a bit more complicated.

The detailed procedure about the derivation of the cut-1
absorptive part of the fermionic
triangle was given in \cite{Arg}, so here we only present the results
corresponding to the $Wt\bar b$ vertex, as
\begin{equation}
\label{new}
(\Delta\Gamma^{\alpha\beta\mu})^{(1)}=C_0^{(1)}p_+^{\alpha}p_+^{\beta}p_+^{\mu}+
C_1^{(1)}q^{\alpha}p_+^{\beta}p_+^{\mu}+C_2^{(1)}g^{\beta\alpha}p_+^{\mu}+
C_3^{(1)}g^{\mu\beta}p_+^{\alpha}+C_4^{(1)}g^{\mu\beta}
q^{\alpha}+C_5^{(1)}g^{\mu\alpha}p_+^{\beta},
\end{equation}
where the superscript (1) denotes the quantities for the cut-1.

Comparing with the corresponding expressions for
$Z^{\alpha\beta\mu}$
given in \cite{Arg}, one can notice that there are two more terms in (\ref{new}).
Moreover, the expressions of $C_0^{(1)}\sim C_5^{(1)}$ are much more complicated
than that of \cite{Arg}, in fact, the derivations are similar, even though
very tedious. The explicit expressions of $C_0^{(1)}\sim C_5^{(2)}$ are
presented in the appendix.

The cut-1 and cut-2 are geometrically symmetric to the virtual photon line
in the triangle diagram, so that while deriving the absorptive part of the loop
corresponding to cut-2, one obtains
$C_0^{(2)}\sim C_5^{(2)}$ by an exchange of $p_+\leftrightarrow p_-$.
However, since $p_-$ is associated with the incoming W-boson, we have
$$\epsilon_{\beta}p_-^{\beta}=0,$$
the terms related to $p_-^{\beta}$ disappear and the Lorentz structure for
cut-2 is exactly the same as that of \cite{Arg} where $Wu\bar d$ vertex was
under consideration.

The total absorptive contribution of the fermionic triangle is
\begin{equation}
\Delta\Gamma^{(tot)}_{\alpha\beta\mu}=\Delta\Gamma^{(1)}_{\alpha\beta\mu}+
\Delta\Gamma^{(2)}_{\alpha\beta\mu}.
\end{equation}
$\Delta\Gamma^{(1)}_{\alpha\beta\mu}$ and
$\Delta\Gamma^{(2)}_{\alpha\beta\mu}$ would be in opposite sign, because of
a fermionic exchange. If they were equal in magnitude, they would cancel each
other. This observation is consistent with eq.(\ref{vio}). Obviously, in fact,
they are not equal in magnitude, and they would cancel $i\gamma_W$ and
$i\gamma_W'$ respectively as discussed above.

Argyres et al. proved in a convincing way that the existence of the absorptive
part of the fermionic triangle fully compensates the effect of $i\gamma_W$
and thus restores the U(1) gauge invariance. Therefore, if one adds $\Delta
\Gamma_{\alpha\beta\mu}$ to the tree expression, the U(1) gauge invariance is
guaranteed and their results indicated that the troublesome term $s/m_e^2$
does not exist in evaluating the cross section.

However, as noted that the expressions of $\Delta\Gamma^{(tot)}_{\alpha\beta
\mu}$ is very complicated and its practical applications are restricted. We
introduced a simple scheme as the "projection scheme" which can eliminate the
linear divergent $s/m_e^2$ automatically.

In the following section, we will discuss its application to $e^-+W^+\rightarrow
e^-t\bar b$ process.\\

(3) The application of the projection scheme.

The projection scheme is described in all details in \cite{Chang}. For
completeness, here we give a brief introduction to the scheme.

In fact, the U(1) gauge invariance i.e. the Ward identity guarantees
an exact cancellation among
unphysical large numbers and only physical quantities which may be many
orders smaller than the large numbers remain. Therefore, even small violation
of the gauge invariance fails the delicate cancellation and
leads to disastrous divergent terms such as $s/m_e^2$. It is well understood
that this gauge invariance violation is due to an ill-truncation of the
perturbative expansion, so is not physical. If we first confirm the gauge
invariance,  we can replace
\begin{equation}
l_{\mu}\equiv \bar u_e(p_2)\gamma_{\mu} u_e(p_1),
\end{equation}
by
\begin{equation}
l'_{\mu}=l_{\mu}-cq_{\mu}.
\end{equation}
Since $q_{\mu}T^{\mu}=0$ where $T_{\mu}$ is the effective vertex of Fig.1, the
replacement is trivial if there is no gauge invariance violation. However,
when $i\gamma_{W}$ exists this treatment plays an crucial role for eliminating
the linear divergence. We choose the coefficient $c$ according to
\begin{equation}
{\delta\over\delta c}max(|l'_0|^2,|l'_1|^2,|l'_2|^2,|l'_3|^2)=0,
\end{equation}
and obtain
\begin{equation}
c={q_0l_0+q_1l_1+q_2l_2+q_3l_3\over q_0^2+q_1^2+q_2^2+q_3^2}
\end{equation}
where $k_i$ are in the Euclidean space \cite{Chang,Zhu}.
Because
$$q_{\mu}l^{\mu}=0,$$
in the Minkovsky space, we have
$$c={2k_0q_0\over ||k||^2},$$
where $||k||^2\equiv k_0^2+k_1^2+k_2^2+k_3^2$ is a positive definite quantity.
This scheme, in principle, is to select a special gauge and it is aimed to
eliminate the components of $l_{\mu}$ which are parallel to $q_{\mu}$ in the
Euclidean space.

By the projection scheme, the problem of large number cancellation is avoided,
the troublesome divergent $s/m_e^2$ does not appear at all. This scheme is
embedded in the program FDC which is designed for calculating such cross
sections \cite{Wang}. With the program FDC, we calculate the differential
and total cross sections of $e^-+W^+\rightarrow e^-+t+\bar b$ for various
energies. The results are tabulated bellow where for a comparison, we list the
numbers corresponding to the results of the standard (i.e. without $i\gamma_W$),
without using the projection scheme and with the scheme respectively.\\

\centerline{Table 1.}

\begin{center}
\vspace{0.3cm}
\begin{tabular}{||c|c|c|c|c|c|c||}
\hline
$\sqrt s$ (GeV) & 200  & 500 & 800 & 1200 & 1600 & 2000 \\
\hline
"standard" & $6.44\times 10^{-2}$ & $6.92\times 10^{-1}$
& 1.04 & 1.30 & 1.46 & 1.57\\
\hline
without PJ & $1.18\times 10^{2}$ & $8.32\times 10^{5}$ &
$7.18\times 10^{6}$ & $2.98\times 10^{7}$ & $1.30\times 10^{8}$ &
$3.21\times 10^8$\\
\hline
with PJ & $6.44\times 10^{-2}$ & $6.92\times 10^{-1}$ &
1.04 & 1.30 & 1.46 & 1.57\\
\hline
\end{tabular}

\vspace{0.3cm}

\end{center}

The notation in the table 1, "standard" means that results obtained without
introducing $i\gamma_W$ in the W-boson propagator and also without employing
the projection scheme; "without PJ" means the results obtained with $i\gamma_W$
but without using the projection scheme and "with PJ" denotes the results with
$i\gamma_W$ and the projection scheme. The unit of the cross section is in
GeV$^{-2}$.
Our results are also graphed in Fig.3 and Fig.4.

In Fig.3, one can see that the differential cross section $d\sigma/d\cos\theta$
tends to infinity as $\theta\rightarrow 0$. That indicates  a collinear
divergent property of the process. It is also noted that with the projection
scheme, even though $d\sigma/d\cos\theta$ is still singular at $\theta=0$,
the degree of singularity is decreased compared to that without projection
scheme. One can expect that this singularity of differential cross section
would not blow up the total cross section.
The curves in Fig.4. confirms this point.

As discussed above, $i\gamma_W$ does not need to be added in this process, so
the corresponding solution without $i\gamma_W$ can serve as the standard, i.e.
the real physical solution. In Fig.4. it corresponds to curve-1. When the
projection scheme is not employed and the Breit-Wigner form of the W-propagator
is taken, we observe that the results shown in curve-2 are 7$\sim$8 orders
larger than the standard. It indicates that such results are not physical.
The curve-3 corresponds to the results where the Breit-Wigner form is taken and
the projection scheme is employed. We notice that the curve-3 coincides with the
standard (curve-1) perfectly, it can also be seen from the data in table 1.
Namely we almost cannot distinguish between them.\\

(4) Discussion.

The U(1) gauge invariance violation brings up difficulties for correctly
evaluating the cross sections. Without carefully handling this problem, all
obtained results are not reliable at all. In principle, one should restore the
U(1) gauge invariance by involving the loop effects. The complete restoration
which includes higher order contributions is difficult, even though possible
\cite{Para}. Thus a practical method which is applicable to theoretically
calculating the cross sections of all processes where a t-channel virtual
photon exists and collinear divergence would cause disastrous consequence, is
favored.

The projection scheme is not aimed to restore the U(1) gauge invariance, but to
eliminate the troublesome $s/m_e^2$, while $\log(s/m_e^2)$ remains and
contributes negligibly to the total cross section.
Our numerical results indicate that even for
$\sqrt s=2000$ GeV, the consistency of the results with the "standard" is
perfectly satisfactory. Therefore, this scheme is applicable for evaluating
cross sections at very high energies.

Similar schemes were proposed \cite{Bh}, compared to theirs, our scheme is
easier in practical applications, but in principle, all schemes are parallel.

Moreover, the same scheme is embedded in the FDC program and applied to
evaluating the cross section of $e^+e^-\rightarrow e^+e^-$. Since this
process serves as a standard reaction to testify the working condition of a
collider and  software, accurate evaluation of its cross section is crucially
important. Application of the projection scheme is proved to give satisfactory
results.

Our conclusion is that the projection scheme is very useful for evaluating
the cross sections where U(1) gauge invariance violation might be brought up
by the Breit-Wigner form of propagators of unstable intermediate bosons or
fermions. The accuracy is satisfactory for a very wide range of energy. This
scheme can be widely applied in the numerical analysis of the future
experiments.\\

\vspace{1cm}

\noindent Appendix\\

\begin{equation}
(\Delta\Gamma^{\alpha\beta\mu})^{(1)}=C_0p_+^{\alpha}p_+^{\beta}p_+^{\mu}+
C_1q^{\alpha}p_+^{\beta}p_+^{\mu}+C_3g^{\mu\beta}p_+^{\alpha}+C_4g^{\mu\beta}
q^{\alpha}+C_5g^{\mu\alpha}p_+^{\beta},
\end{equation}
where
\begin{eqnarray}
C_0 &=& C_{00}+C_{01}f_0, \nonumber\\
C_1 &=& C_{10}+C_{11}f_0, \nonumber\\
C_2 &=& C_{20}+C_{21}f_0, \nonumber\\
C_3 &=& C_{30}+C_{31}f_0, \nonumber\\
C_4 &=& C_{40}+C_{41}f_0, \nonumber\\
C_5 &=& C_{50}+C_{51}f_0.
\end{eqnarray}
The coefficients $C_{i0}$, $C_{i1}$ and the function $F_0$ are given below
explicitly.
\begin{eqnarray}
C_{00} &=& {1\over\lambda}({-40\over 3}P_+\cdot q-{52\over 3}p_+^2+20 q^2)
\nonumber\\
&& +{1\over\lambda^2}({152\over 3}p_+\cdot qp_+^2q^2-{26\over 3}p_+^4q^2
-4p_+\cdot q q^4-18p_+^2q^4)\nonumber \\
&& +{1\over\lambda^3}(-30p_+\cdot qp_+^4q^4+10p_+^6q^4-10p_+\cdot qp_+^2q^6
+30p_+^4q^6);\\
C_{01} &=& -4+{1\over\lambda}(4p_+\cdot qp_+^2+8p_+\cdot qq^2-10p_+^2q^2
-6q^4) \nonumber\\
&& +{1\over\lambda^2}(3p_+\cdot qp_+^4q^2-9p_+\cdot qp_+^2q^4-9p_+^4q^4
+2p_+\cdot qq^6-7p_+^2q^6)\nonumber\\
&& +{1\over\lambda^3}(5p_+\cdot qp_+^6q^4+30p_+\cdot qp_+^4q^6-20p_+^6q66
\nonumber\\
&& +5p_+\cdot qp_+^2q^8-20p_+^4q^8);\\
C_{10} &=& {1\over\lambda}({-100\over 3}p_+^2) \nonumber\\
&& +{1\over\lambda^2}({-4\over 3}p_+\cdot qp_+^4+20p_+\cdot qp_+^2q^2-{20\over 3}
p_+^4q^2+8p_+^2q^4) \nonumber\\
&& +{1\over\lambda^3}(-10p_+\cdot qp_+^6q^2-30p_+\cdot qp_+^4q^4+30p_+^6q^4
+10p_+^4q^6);\\
C_{11} &=& {1\over\lambda}(8p_+\cdot qp_+^2-6p_+^4+8p_+^2q^2)\nonumber\\
&& +{1\over\lambda}(12p_+\cdot qp_+^4q^2-10p_+^6q^2-18p_+^4q^4-4p_+^2q^6)
\nonumber\\
&& +{1\over\lambda^3}(20p_+\cdot qp_+^6q^4-5p_+^8q^4+20 p_+\cdot qp_+^4q^6
-30p_+^6q^6-5p_+^4q^8);\\
C_{20} &=& {1\over\lambda}({-8\over 3}p_+\cdot qp_+^2+6p_+^2q^2) \nonumber\\
&& +{1\over\lambda^2}
(-2p_+\cdot qp_+^4q^2-2p_+\cdot qp_+^2q^4+4p_+^4q^4);\\
C_{21} &=& -2p_+\cdot q+2p_+^2+{1\over\lambda}(-2p_+\cdot qp_+^2q^2) \nonumber\\
&& +{1\over\lambda^2}(3p_+\cdot qp_+^4q^4-p_+^6q^4+p_+\cdot qp_+^2q^6-
3p_+^4q^6);\\
C_{30} &=& 4+{1\over\lambda}({16\over 3}p_+\cdot qp_+^2+4p_+\cdot qq^2-
6p_+^2q^2)\nonumber\\
&& +{1\over\lambda^2}(-2p_+\cdot qp_+^4q^2-2p_+\cdot qp_+^2q^4+4p_+^4q^4);\\
C_{31} &=& -6p_+\cdot q-2p_+^2+8q^2\nonumber\\
&& +{1\over\lambda}(-2p_+\cdot qp_+^2q^2-p_+^4q^2-2p_+\cdot qq^4+3p_+^2q^4)
\nonumber\\
&& +{1\over\lambda^2}(3p_+\cdot qp_+^4q^4-p_+^6q^4+p_+\cdot qp_+^2q^6
-3p_+^4q^6);\\
C_{40} &=& {1\over\lambda}(4p_+\cdot qp_+^2+{2\over 3}p_+^4-4p_+^2q^2)
\nonumber\\
&& +{1\over\lambda^2}(-4p_+\cdot qp_+^4q^2+2p_+^6q^2+2p_+^4q^4);\\
C_{41} &=& 2p_+^2+{1\over\lambda}(2p_+\cdot qp_+^4-4p_+\cdot qp_+^2q^2
-2p_+^4q^2+2p_+^2q^4)\nonumber\\
&& +{1\over\lambda^2}(p_+\cdot qp_+^6q^2+3p_+\cdot qp_+^4q^4-3p_+^6q^4
-p_+^4q^6);\\
C_{50} &=& {1\over\lambda}({40\over 3}p_+\cdot qp_+^2+{2\over 3}p_+^4-
10p_+^2q^2)\nonumber\\
&& +{1\over\lambda^2}(-6p_+\cdot qp_+^4q^2+2p_+^6q^2-2p_+\cdot qp_+^2q^4
+6p_+^4q^4);\\
C_{51} &=& 2p_+\cdot q-4p_+^2\nonumber\\
&& +{1\over\lambda}(2p_+\cdot qp_+^4+2p_+\cdot qp_+^2q^2-6p_+^4q^2-2p_+^2q^4)
\nonumber\\
&& +{1\over\lambda^2}(p_+\cdot qp_+^6q^2+6p_+\cdot qp_+^4q^4-4p_+^6q^4
+p_+\cdot qp_+^2q^6-4p_+^4q^6).
\end{eqnarray}

We also present
\begin{equation}
f_0={1\over\sqrt{\lambda}}\log{p_-\cdot q+\sqrt{\lambda}\over
p_-\cdot q-\sqrt{\lambda}},
\end{equation}
and
$$\lambda=(p_+\cdot q)^2-p_+^2q^2.$$

\vspace{2cm}
Figure Caption\\
Fig1. The Feynman diagrams at tree level for $e^-+W^+\rightarrow e^-+t+\bar b$.\\

Fig.2. The fermionic triangle. The cut-1 and cut-2 are drawn and they correspond
to different absorptive parts of the loop.\\

Fig.3. The differential cross section $d\sigma/d\cos\theta$ vs. $\theta$
with $\sqrt s=1000$ GeV, the curve-1 stands for the standard, the curve-2 for
the results without PJ and the curve-2 for that with PJ.\\

Fig.4. The total cross sections $\sigma$ vs. $\sqrt s$. The curve-1 stands for
the standard, the curve-2 for the results without PJ and the curve-3 for
that with PJ.

\end{document}